\documentstyle[12pt]{article}
\input epsf

\begin{document}

\titlepage

\def\expt#1{\mathord{<}#1\mathord{>}}
\def\Yb{\bar Y}
\def\Ub{\bar U}
\def\Tb{\bar T}
\def\Sb{\bar S}
\def\Wb{\bar W}
\def\e{\epsilon}
\def\beq{\begin{equation}}
\def\eeq{\end{equation}}
\def\beqa{\begin{eqnarray}}
\def\eeqa{\end{eqnarray}}
\def\be{\begin{equation}}
\def\ee{\end{equation}}

\def\bi{\begin{itemize}}
\def\ei{\end{itemize}}
\def\i{\item}

\begin{flushright}
  TUM-HEP-211/95\\
   UPR-653-T\\
   hep-th/9503170\\
\end{flushright} \vspace{1ex}

\begin{center} \bf

  GAUGINO CONDENSATION, S-DUALITY \\ AND \\ SUPERSYMMETRY BREAKING \\ IN
  SUPERGRAVITY MODELS \\ \rm \vspace{1ex}

{\bf Z. Lalak$^{\dagger,\ast}$, A. Niemeyer$^\dagger$ and
H. P. Nilles$^{\dagger,\ddagger}$}

\vspace{0.3cm}
 $^\dagger$ {Physik Department} \\
       {\em Technische Universit\"at M\"unchen} \\
       {\em D--85748 Garching, Germany}

\vspace{0.3cm}
 $^\ddagger$ Max Planck Institut f\"ur Physik \\
  {\em Heisenberg Institut}\\
  {\em D--80805 Munich, Germany}

\vspace{0.3cm}
 $^\ast$  University of Pennsylvania\\
  {\em Physics Department}\\
  {\em Philadelphia PA 19104-6396}
\vspace{0.5cm}

ABSTRACT
\end{center} The status of the gaugino condensation as the source of
supersymmetry breaking is reexamined.  It is argued that one cannot have stable
minima with broken supersymmetry in models where the dilaton is coupled only
linearly
to the gaugino condensate.  We show that the problems of the gaugino condensate
mechanism can be solved by considering nonstandard gauge kinetic
functions, created by nonperturbative effects. As an example we use
the principle of S-duality to modify the coupling of the gaugino condensate
to effective supergravity (superstring) Lagrangians. We show that such an
approach can solve
the problem of the runaway dilaton and lead to satisfactory supersymmetry
breaking in models with a {\em single} gaugino condensate.
We exhibit a general property of theories containing a symmetry acting on the
dilaton and also shed some light on the question whether it
is generically the auxiliary field of the modulus $T$, which dominates
supersymmetry breaking.

\newpage

\section{Introduction}

Understanding the ways supersymmetry can possibly be broken in
supersymmetric theories unifying fundamental forces is likely to be the most
challenging problem in supersymmetric model building over the coming years.
While
the agreement that low energy supersymmetry has to be broken on a TeV scale to
allow for realistic phenomenology is unanimous, the hunt for a technically
satisfactory and esthetically compelling model of supersymmetry breaking
continues.  It has long been concluded that spontaneous breaking of
supersymmetry in the observable sector of supersymmetric models cannot work
properly, because of unbroken supersymmetric mass relations between known
particles and their superpartners (vanishing supertrace) which practically make
superpartners visible in some way in all the models analyzed so far \cite{rev}.
This
makes it necessary to construct hidden sectors in these models, where
supersymmetry can be broken spontaneously and the effect of this breaking can
then be transmitted by sufficiently weak couplings to the observable sector --
where it is seen as explicit (soft) breaking. The problem with
this scenario consists in the necessity of providing an explicit (and rather
large
if the hidden sector is to decouple at low energies) mass scale which can be
used to generate a TeV scale in the observable sector. The only known
way of generating such an intermediate scale in field theoretical
models is to employ nonabelian Yang-Mills theories, the condensation scale of
which can easily account for the required order of magnitude of the new scale
\cite{nil1}, \cite{nfer}.
In this case the source of really weak couplings is gravity. Thus the
nonabelian gauge sector of a superstring model is a natural hidden sector based
on well known physics.
This path has been explored by many authors over the years
\cite{nil1}-\cite{nrev}, but unfortunately no fully satisfactory solution for
realistic
SUSY breakdown in supergravity, and in particular in superstring inspired
effective supergravities has been found. The main flaw of the models considered
in the literature is the difficulty of fixing the the vacuum expectation value
of the dilaton, the field which sets the value of the gauge coupling at the
unification scale -- so crucial for the dynamics of the gauge sector -- at a
physically acceptable value.  Solutions proposed so far in the context of
gaugino condensation involve several gaugino condensates and a rather
unnatural adjustment of the hidden matter sector \cite{ccm}.  Of course, it may
be that because of some unknown and deep reason string theory will generate
these special matter sectors with suitably designed superpotentials and
couplings. However, at the level of field theory, and gaugino condensation in
itself is inherently a field theoretical phenomenon, one would prefer that if
condensates break supersymmetry they should do it in a generic way,
irrespectively of accidental complications of the matter sector. Further to
that, if there is an obvious obstacle for the mechanism to work, as we believe
the form of perturbative couplings of the dilaton in stringy Lagrangians to be,
its possible absence should be understandable in terms of some fundamental
symmetry.

In this work we want to analyze as far as possible the ways the pure
gauge  hidden sector, coupled eventually to the dilaton and other moduli,
can break supersymmetry and identify crucial problems. As the mechanism
of the formation of condensates is nonperturbative, it is not obvious what
technical tools should be used for such an analysis. As we explain in more
detail in the text, we decided to use the generalized effective Lagrangian
approach, which offers a hope for controlling a variety of dynamics pertaining
to the condensation in the presence of a possible back-reaction of gravity,
moduli
fields and other fields typically present in realistic models.
Indeed, it turns out that the role the gaugino condensate can play is rather
model dependent. Moreover, the presence and nature of the couplings of dilaton
and other moduli to condensates and to gravity appears to be crucial for
the supersymmetry breaking with simultaneous stabilization of all the moduli
to work. In fact, on the basis of our analysis we tend to argue that
the linear coupling of the dilaton to the gaugino condensate --- motivated by
string tree level amplitudes \cite{wit} --- and therefore the effective
superpotential for
the dilaton in its customary form, is highly unsatisfactory and that the
dilaton-induced
problems we discuss provide an evidence for the need of a more sophisticated
structure of the strongly coupled sectors of supersymmetric gauge theories.
We suggest that it is  possible to solve  the problem of the
runaway dilaton/SUSY breaking  in a  fundamental way, for example by
postulating  a new symmetry of  the
effective  Lagrangian, the so-called S-duality. It turns out that the
modification of dilaton couplings induced by even the simplest versions of
S-duality is sufficient
to stabilize the dilaton at a reasonable value and allows for
supersymmetry breaking.
Evidence for S-duality in string models has been presented in \cite{witsen},
and it might be that S-duality gives the proper way of promoting the inherently
field-theoretical gaugino condensation mechanism into
the string theoretical framework.

S-duality invariant effective purely dilatonic superpotentials have been
conjectured in \cite{fi} (however with no reference to gaugino condensation)
and recently reexamined by the authors of \cite{hm}. The general form of the
superpotential proposed in \cite{fi}, which was constructed specifically to fix
the vev of the dilaton, does not give a free theory in the weak
coupling limit. The authors of Ref.\ \cite{hm} note
that one can easily modify any effective
superpotential in such a way, that it vanishes asymptotically as $\mbox{Re} S
\rightarrow \infty$ in any direction. Their one-condensate model
shows a realistic minimum for the dilaton but, unfortunately, SUSY is unbroken
at this minimum.

Our paper is structured as follows: in chapter 2 we discuss
the effective Lagrangians for gaugino condensate in globally supersymmetric
sigma models containing condensates and the dilaton. In chapter 3 we
couple these models to supergravity. In chapter 4 we introduce S-dual
effective Lagrangians, in chapter 5 we discuss general aspects of
supersymmetry breaking in the class of models of interest, and finally
summarize our results in chapter 6.

\section{Effective Lagrangian in global SUSY-Yang-Mills theory}

Following the pioneering work of \cite{veny} we discuss in some detail the
construction of the effective Lagrangian which is supposed to describe
gaugino condensation in the globally supersymmetric Yang-Mills
theory. The underlying principle of this construction is t'Hooft's anomaly
matching condition, which demands that the effective low energy Lagrangian,
valid below some scale $\Lambda$ (which is to be identified with the
condensation scale in our case), should reproduce the anomalies of the
underlying constituent theory. In the case of the Yang-Mills model it is well
known \cite{ferzum} that R-symmetry and supersymmetry currents as well as the
energy momentum tensor lie in the same (general) supersymmetry multiplet.
Moreover, this implies that the corresponding chiral, $\gamma$-trace and trace
anomalies lie in another, chiral, multiplet which we will denote by $U$. For
the
sake of convenience we will sometimes use the chiral multiplet $Y$, defined by
$U=Y^3$, which has canonical mass dimension.  In terms of constituent fields
the lowest component of $U$ is proportional to the gaugino bilinear $\lambda
\lambda$, so it makes sense to take $U$ as the (pseudo-)goldstone multiplet
entering the low energy Lagrangian. When we define components of $U$ as $U=A +
\sqrt{2} \theta \psi + \theta \theta F$ then the R-symmetry acts as

\beq
A \rightarrow e^{3 i \alpha} A, \; \psi \rightarrow e^{3 i \alpha /2} \psi,\;
F \rightarrow F
\eeq

which can be also written as

\beq
U(x,\theta) \rightarrow e^{3 i \alpha} U(x, e^{-3 i \alpha /2} \theta)
\label{er}
\eeq

while the scale symmetry acts as

\beq
A(x) \rightarrow e^{3 \gamma} A(x e^{\gamma}), \; \psi(x) \rightarrow
e^{7 \gamma /2} \psi(x e^{\gamma}), \; F(x) \rightarrow e^{4 \gamma} F(x
e^{\gamma})
\eeq

which one can write more concisely

\beq
U \rightarrow e^{3 \gamma} U(x e^{\gamma}, \theta e^{\gamma /2})
\label{scal}
\eeq

Assuming that the anomalies associated with the above classical invariances
are reproduced by the superpotential W, one obtains the (holomorphic)
equation for the superpotential

\beq U \frac{\partial W}{\partial U} -W = b U, \eeq

which has a general solution of the form

\beq
W = a \; U + b \; U \log(\frac{U}{\mu^3}),
\label{supg}
\eeq

where $a$ is in general undetermined - as the variation of $F$ (highest
component of $U$) under
both symmetries vanishes - and corresponds to the rescaling of the
condensation scale $\mu$, and $b$ is some function of the gauge coupling which
is fixed by demanding the specific form of the resulting anomaly coefficient.
To have the complete model one has to make a choice for the K\"{a}hler
potential  $K$. Usually one demands that the variation of $K_D$
(D-component of $K$) is
non-anomalous which fixes it, up to a field-independent coefficient, to be
$K=9 (U \bar{U})^{1/3}=9 Y\bar Y$. This choice leads to unbroken supersymmetry
and  nonzero condensate with the expectation value $U=\mu^3 e^{-(a+b)/b}$.
This agrees with the conclusion of
reference \cite{witt} where the analysis based on the index theorem
implies that supersymmetry is unbroken in this case, and with instanton
calculations \cite{rossi}. However, these methods do not give any
information about the form of the kinetic term for the condensate, and in
our present approach the conclusion that supersymmetry stays unbroken
is almost independent of the form of $K$, no matter whether it
breaks classical invariances or not.
Indeed, the effective potential, for any $K$, has the form

\beq
V=(K_{U \bar{U}})^{-1} |U|^2 |a+b \log(\frac{U}{\mu^3})|^2,
\label{eft}
\eeq

($K_{U\Ub} = \partial^2 K/\partial U \partial \Ub)$ and the expression
controlling supersymmetry breaking is

\beq
F^U =(K_{U \bar{U}})^{-1} \bar{U} (a+b \log(\frac{\bar{U}}{\mu^3})),
\eeq

where $F^U$ denotes the auxiliary field of the condensate superfield $U$.

One can see that $V=0$ implies $F=0$, regardless of the particular form of $K$
(unless the metric $K_{U \bar{U}}^{-1}$ becomes singular). It should be noted,
that the demand that SUSY is unbroken does not fix the form of $K$ at all, does
not even imply that $K$ is invariant under classical invariances.
We will see however, that the choice of $K$ is not an innocent assumption if
one considers supergravity models. It should also be noted, that there is no
canonical form of $K$ implied by any reliable calculation.

It is instructive to examine in more detail the variations of $K_D$ under
chiral and scale transformations.  The infinitesimal change under chiral
symmetry (\ref{er}) is \beq \delta K_D = 3 i \alpha (\frac{\partial K}{\partial
  U} U - \frac{\partial K}{\partial \bar{U}} \bar{U})_D
\label{ver}
\eeq

and under scale symmetry (\ref{scal})

\beq \delta (\int d^4 x K_D ) = \gamma \int d^4 x ( 3 \frac{\partial
  K}{\partial U} U + 3 \frac{\partial K}{\partial \bar{U}} \bar{U} -2 K)_D
\label{vscal}
\eeq

Let us consider two simple examples. If one takes $K=K(U \bar{U})$
then the variation (\ref{ver}) vanishes identically, but
(\ref{vscal}) is nonzero unless $K = const \; (U \bar{U})^{1/3}$.
If one considers $K=K(U+\bar{U})$ this gives a non-invariant $K_D$
under chiral transformation, but the special choice $K=(U+\bar{U})^{2/3}$
is invariant under the scale transformation (\ref{scal}).

Finally, let us note that for the choice $K=c \; U \bar{U}$, which will be
discussed in the next section, the
non-vanishing variation (\ref{vscal}) is
\beq
\delta_{\gamma} (\int d^4 x (U \bar{U}) = 4 \gamma ( |F^U|^2 + \;
fermionic\; \; and \; \; derivative \; \; terms \; )
\eeq
which vanishes if supersymmetry is unbroken (which is the case for this
choice of $K$ in global supersymmetry).

The natural generalization of the above Lagrangian consists in allowing
for a dynamical, field-dependent gauge coupling. As the global Yang-Mills
Lagrangian contains a  term $(1/g^2 \; W^{\alpha} W_{\alpha})_F
+ h.c.$ then it is natural to promote the inverse gauge coupling constant
to a chiral superfield, f. This extension is well motivated by
the superstring effective Lagrangian which gives $f=S$ at the string
tree-level \cite{wit}, where $S$ denotes  the dilaton, and
also predicts a characteristic
no-scale type K\"ahler function for the dilaton, $K=-\log (S+\bar{S})$.
Thus the modified Lagrangian, which can be considered as the flat global
limit of some  supergravity model dilaton plus gaugino condensate sector
is

\beq
L=K(S,\bar{S})_D + K(U,\bar{U})_D + (( f(S) U + b U \log (\frac{U}{\mu^3}))_F +
h.c.\;)
\eeq

We assume, as it is the case in superstring models, that the dilaton
$S$ is dimensionless, which in the context of supergravity implies
that the associated dimensionful field is $\hat{S} = S \; M$ where
$M=M_{pl}$.
Formally $M$ can be an arbitrary scale, but it is natural to assume that there
are no fundamental scales in the model other than $M_{pl}$ and the condensation
scale.
The introduction of the dilaton, which is a gauge singlet, requires that we
extend both chiral and scale symmetries to act on the dilaton in such a way
that under R-symmetry $f(x,\theta) \rightarrow f(x, e^{-3i\alpha/2} \theta)$
and under scale transformation $f(x,\theta) \rightarrow f(x e^{\gamma/2},\theta
e^{\gamma /2})$.
In addition it creates an anomalous symmetry which
is the global shift of the imaginary part of $f$, $f \rightarrow f+i\Lambda$.
This symmetry is an exact symmetry of the corresponding sigma model if
$K=K(f+\bar{f})$. Because of the presence of this new shift symmetry
and because of the holomorphicity of the superpotential there appear two
new exact symmetries of the superpotential. The first, which is generally
assumed to be the exact symmetry of the full Lagrangian through the suitable
choice of
$K(S,\bar{S})$, is the combination of the R-symmetry (\ref{er}) and the
imaginary shift

\beq
f(x,\theta) \rightarrow f(x,e^{-3i\alpha/2})- 3ib \alpha
\label{rsh}
\eeq

The second symmetry, which is violated by usual (for instance, the
 no-scale form) K\"ahler functions, is (\ref{scal}) accompanied by

\beq
f(x,\theta) \rightarrow f(x e^{\gamma} ,e^{\gamma/2})- 3 b \gamma
\label{gsh}
\eeq

Even though the second symmetry is not a symmetry of the kinetic terms,
its existence implies an additional degeneracy of the manifold  of
supersymmetric solutions of the effective model \cite{wesov}.

Let us discuss implications of the presence of the dilaton for supersymmetry
breaking issue. The general potential we consider has the form

\beq V = g^{U\bar{U}} |f+b \log (\frac{U}{\mu^3})|^2 +g^{S \bar{S}} |U|^2
|\frac{\partial f}{\partial S}|^2
\label{pots}
\eeq

where $g^{i \bar{j}} = (\partial^2 K / \partial z^i \partial z^{\bar{j}})^{-1}$
is the inverse K\"ahler metric. If we assume the most symmetric choices for the
K\"ahler potential, $K=-\log(S+\bar{S}) +9 (U \bar{U})^{1/3}$ and $f=S$, then
the minimal value of (\ref{pots}) corresponds to $U=0$ with any value of $S$
which also gives unbroken supersymmetry. Thus the introduction of the dilaton
not only prevents supersymmetry breaking, but also `undoes'
condensation - one is forced to conclude that the condensate does not form in
this
case.  The general form of (\ref{pots}) suggests however some possibilities.
First, one can imagine that $f$ has such a form, that the second term in
(\ref{pots}) cancels without driving $U$ to zero, fixing at least partially the
vev of the dilaton. Then the terms under the sign of the absolute value in
the first term adjust themselves to cancel this term as well. In this case
supersymmetry is generically unbroken in global models, but at
least one can save the condensate creating a new manifold of (supersymmetric)
vacua, which is disconnected from $U=0$. The second, interesting, possibility
is, at the level of this global model, that one takes a non-symmetric K\"ahler
function for $U$, for instance $K=U \bar{U}/ \mu^4$. Then the value of $U=0$
does not automatically cancel the first term in the potential. The only
supersymmetric point in the hyperplane $U-S$ is the asymptotic minimum of the
potential at $U \rightarrow 0$ and $S \rightarrow \infty$. At any finite value
of S and U supersymmetry is broken and potential is larger than zero. This
example suggests a reasonable way to solve the problem of generating SUSY
breaking at finite values of both dilaton and condensate in purely
dilaton-condensate models. The observation is that there may exist corrections
to the global model which prevent the dilaton from running to infinity, but
stop it at the Planck scale (in our normalization with $M=M_{pl}$ around $S=1$)
instead -- corrections coming from gravity. In that case one expects the
potential (\ref{pots}) to be the leading term in the full potential and
supersymmetry to be broken through the auxiliary field of the dilaton (the
$F^U$-term adjusts itself to zero by means of small corrections when S changes)
with the characteristic value $F^S \approx \mu^3 /M_{pl}$. Unfortunately, we do
not know how to construct a model which works this way, at least without
introducing any mass scale different from Planck and condensation scales into
the dilatonic superpotential $f$.  However, the idea that gravitational
corrections can modify the effective Lagrangian in an interesting way is
carefully discussed in the next chapters where we couple the dilaton-condensate
system to supergravity multiplet.

\section{Gaugino condensate coupled to supergravity}

There are three different ways of introducing gaugino condensates into
supergravity.  In the component Lagrangian method, pioneered by \cite{nfer},
one
takes the standard Lagrangian of supersymmetric Yang-Mills theory coupled to
supergravity and after identifying Lorentz-invariant gaugino bilinears one
replaces them by a constant of the order of the condensation scale $\mu^3$.
Such a procedure has the drawback of discarding the backreaction of other
fields in the model onto the condensate, hence one cannot determine this way
whether the condensate really forms. The formation of the condensate and its
magnitude is simply assumed in this approach implicitly relying on the
observation made in the global version of the model. However, as we assume that
gravitational corrections can play an important part, the internal consistency
of such an approach is not entirely clear. Furthermore, to arrange for broken
SUSY one needs a nontrivial gauge kinetic function, and then one has to check
whether condensates of other fields present in the model do not cancel the
contribution of the assumed gaugino condensate to the vacuum expectation values
of the auxiliary fields.  A refinement of this method which takes into account
a possible dependence of condensate on some other fields (like the dilaton)
leads to the effective superpotential method: here one makes educated guesses
about how the condensate depends on the remaining chiral fields in the model.
Using these one can then construct the gaugino induced ``nonperturbative''
corrections to the original superpotential of the model. For instance, the
belief that the condensate dissolves in the weak coupling limit leads to a
superpotential which decays exponentially with the increasing value of the
dilaton $S$ (large $S$ corresponds to weak coupling) in string inspired
supergravities. One then searches for minima of the effective theory and
determines whether supersymmetry is broken at these minima. The third method is
the effective Lagrangian (Veneziano-Yankielowicz) approach.  Generalizing the
global-SUSY Veneziano-Yankielowicz type Lagrangians discussed in chapter 2, one
should notice that scale invariance may be broken in some sector of the model,
in fact gravity itself introduces an explicit mass scale $M_{pl}$, so that
non-invariant corrections to the K\"ahler potential of the gaugino condensate
become more natural, and that in the superpotential both terms can get
independent chiral-superfield coefficients. Also, one can add an arbitrary
constant to the superpotential (introducing the M-term) which formally does not
affect the global limit.  The standard practical consistency condition is that
in the naive limit $M_{Pl} \rightarrow \infty$, all fields except the
condensate
frozen, one should recover the V-Y Lagrangian. However, in view of the above
arguments, this condition appears to be too strong, as the models whose naive
global limit we take are usually not the pure Yang-Mills models. So, we only
demand that in this naive global limit the supersymmetry breaking is not
induced
by the nonabelian Yang-Mills sector, which is the common conclusion of all
different methods used to analyze global Yang-Mills theories: effective
Lagrangian, index theorem and instanton calculus.  The method we follow here is
the effective Lagrangian approach, as it is best suited to take into account
the dynamics leading to condensation and, perhaps, to supersymmetry breaking.

In supergravity models the effective scalar potential which one has to minimize
is

\beq V= g_{i \bar{j}} F^i F^{\bar{j}} -3 e^G
\label{potloc}
\eeq

where $G=K+\log( |W|^2)$, $ g_{i \bar{j}}=\partial^2 K / \partial z^i
\partial z^{\bar{j}}$ and $F^i$ whose vacuum expectation value signals
supersymmetry breaking is defined to be $F^i = g^{i \bar{j}} \partial G /
\partial z^{\bar{j}} e^{G/2}$.  The simplest model one can consider contains a
condensate with canonical kinetic term (in the following part we will use the
field $Y$ instead of $U$, so that the symmetric kinetic term looks canonical
\cite{venmag})

\be K=Y \bar Y \label{k1}
\ee

and a simple Veneziano-Yankielowicz type superpotential

\be
W=Y^3 (3 \ln \frac Y\mu -1),
\ee

where $\mu$ is the scale at which we expect the condensate to form (whenever
we perform numerical calculations we choose
it to be $\mu=10^{-5} M_{Pl}$).
After minimization of the potential (\ref{potloc}) one discovers that there
are two minima, one at $Y=0$ (vanishing condensate) and one at $Y\neq 0$.
One can easily see, looking at the values of the $F^Y$ term, that at both
minima supersymmetry is unbroken.
This agrees at first glance with the result of the explicit one-instanton
calculation of \cite{tayrey}, where it is argued that instanton effects
do not introduce SUSY breaking even in the local case.
However, one should note that this instanton calculation cannot take
fully into
account the dynamics encoded in the choice of the K\"ahler function K.
In fact, the discrepancy with the one-instanton-induced result reported in
\cite{tayrey} arises even without changing the K\"ahler function, as shown in
the example discussed below.

The easiest way to extend the above toy model consists in the incorporation of
a constant term in the superpotential, which can be thought of as
parameterizing unknown and condensate unrelated effects:

\be W=Y^3 (3 \ln \frac Y\mu -1)+c \label{w1} \ee

For values of the constant $c<1.752 \mu M_{pl}^2$ supersymmetry
remains  unbroken but
if the constant exceeds this value, supersymmetry breaking occurs. The breaking
scale can be adjusted to any value by choosing an appropriate $c$. Fig. 1 shows
the dependence of the expectation value of the auxiliary field $F_Y$ at the
minimum with respect to $c$. The apparent non-analyticity in this plot is not a
numerical artifact as can be seen by looking at the shape of the potentials at
the respective values (Fig. 2). At the critical value for $c$ the minimum with
non-vanishing condensate ceases to exist and reappears again for larger values
of the constant. This shows that some sort of `phase transition' occurs.

\epsfbox[-60 0 500 260]{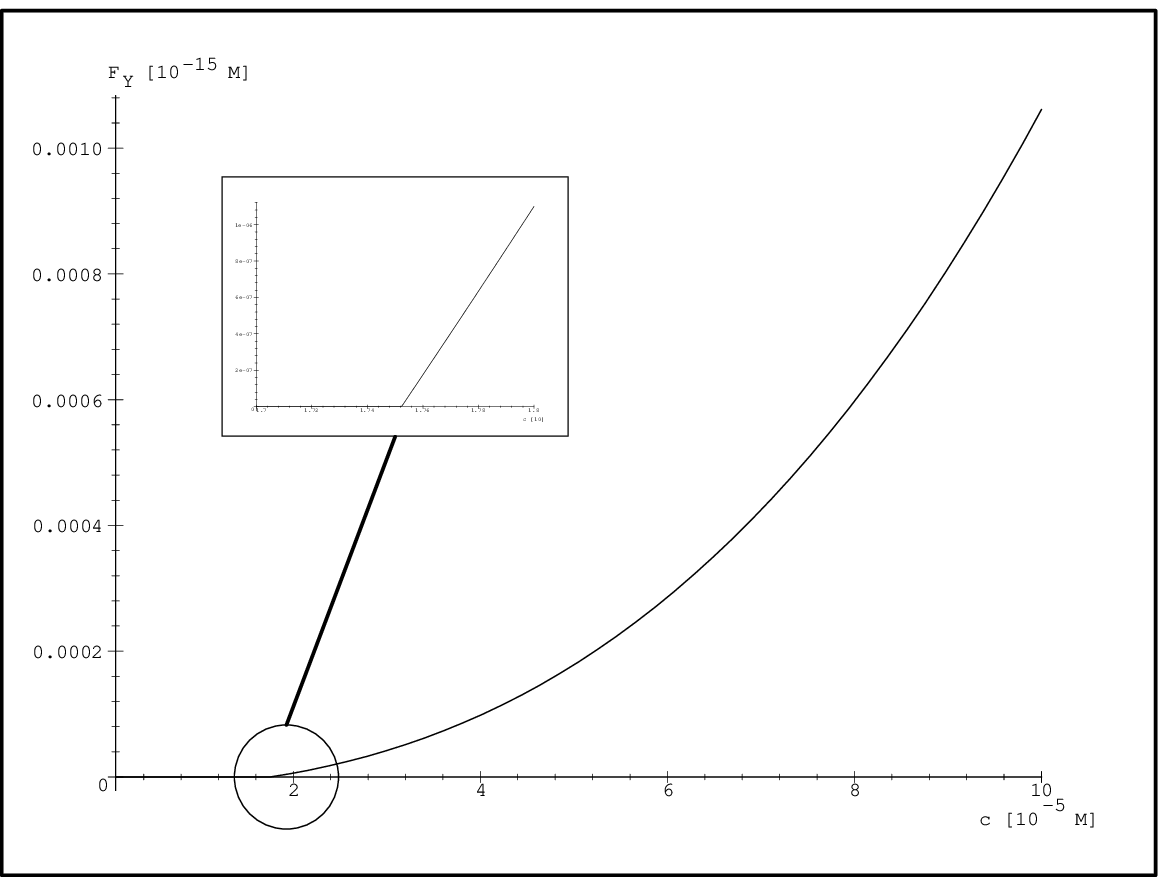}

{\small \em Fig. 1 - Scale of supersymmetry breakdown with respect to the
constant $c$}

\vspace{0.3cm}

\epsfbox[-60 0 500 260]{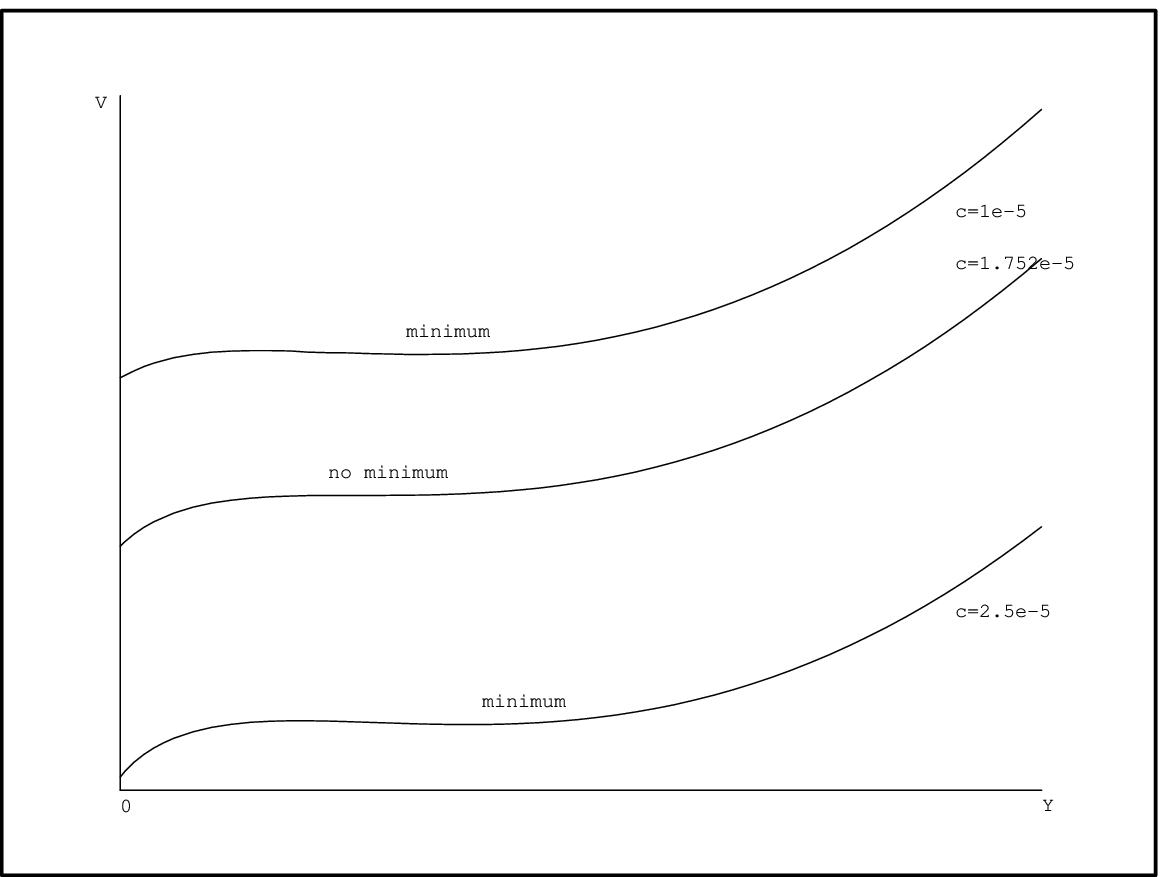}

{\small \em Fig. 2 - Shape of potential for different values of $c$}

\vspace{0.3cm}

The next possibility of extending the basic  model is to take non-minimal
kinetic terms for the gaugino condensate. As proposed by \cite{nil1} one could
take

\be W=Y^3 (3 \ln \frac Y\mu -1) \label{w2} \ee

together with

\be K=3 \ln \frac 3{1-\frac 1c Y^3 \Yb^3/\mu^4 - Y \Yb}\label{k2}, \ee

where the expansion with respect to $Y\Yb$ gives minimal kinetic terms
at the first order.
The general structure of the scalar potential of this class of models
parameterized by $c$ can be described as follows: there exists a minimum with
 $V=0$ at $Y=0$,
corresponding to a vacuum with vanishing condensate. Supersymmetry is unbroken,
of course. For any generic value of $c$ there is also a minimum next to a pole
at nonzero $Y$ (Fig. 3). The cosmological constant of this vacuum may be
positive or negative depending on $c$ and for $c=9$ becomes 0 \cite{nil1}.
Whether supersymmetry is broken at this minimum depends on the value of $c$, as
in the aforementioned model. This model  exhibits a phase transition as well,
which is much
more pronounced. For $c>9+\e$, where $\e \simeq 10^{-6}$, supersymmetry is
unbroken, for $c$ smaller than this value, SUSY is broken (meaning that also
the zero cosmological constant version exhibits SUSY breaking). The change of
the supersymmetry breaking scale is not smooth, but discontinuous. Fig. 4 shows
a plot of the SUSY breaking scale with respect to $x$, with $c=9+x$.
When one considers the global limit of this class of models,
SUSY is always unbroken, with the exception of the value $c=9$.
For this special value of $c$ SUSY is broken in the global limit, and the
metric $g_{Y \bar{Y}}$ vanishes at this point, which means that small
fluctuations of condensate around the minimum do not propagate in the flat
limit.
As the gaugino condensate is the only field considered in these models, it is
of course a trivial statement, that it is the F-term of the condensate
which is responsible for SUSY breaking. One should note, that with
$Y$ being the expectation value of a composite field, also the goldstino is
composite here.

\epsfbox[-60 0 500 260]{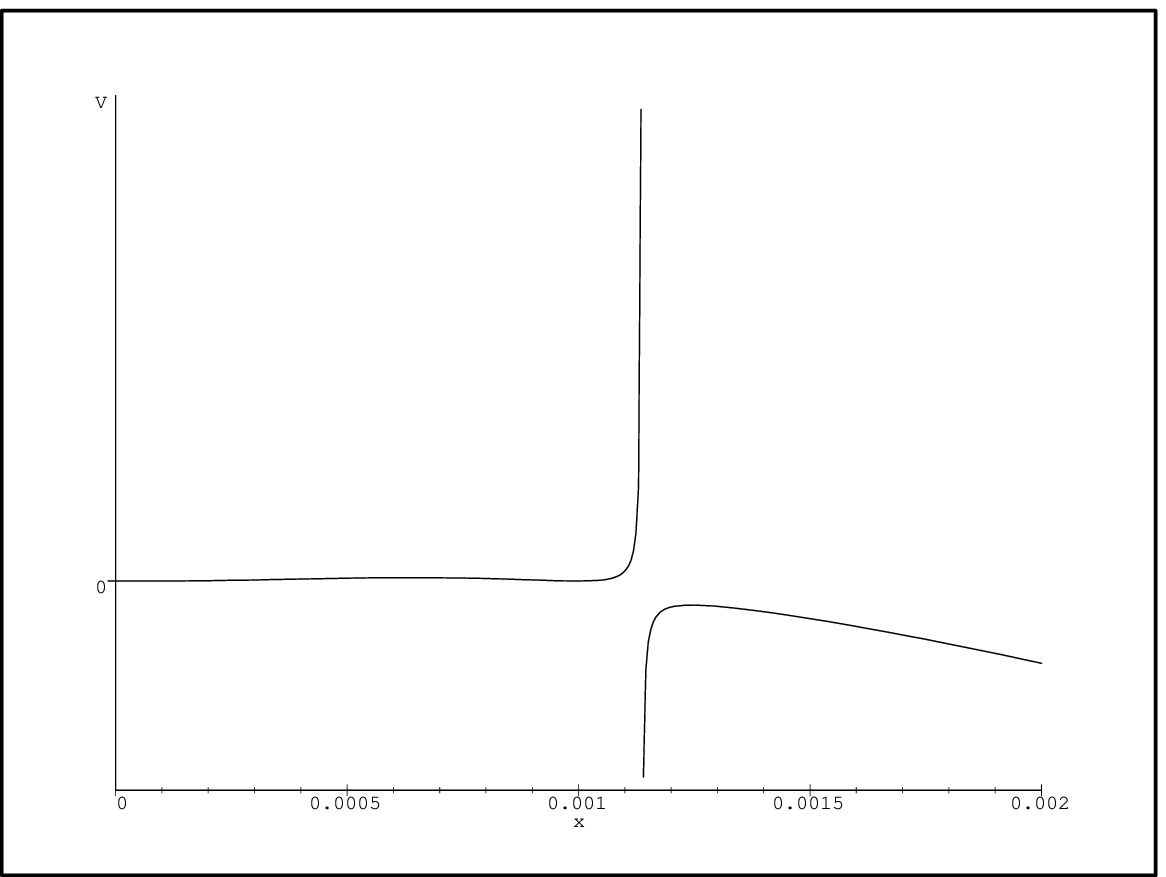}

{\small \em Fig.3 - Shape of $V$ for $c=15$}

\vspace{0.3cm}

\epsfbox[-60 0 500 260]{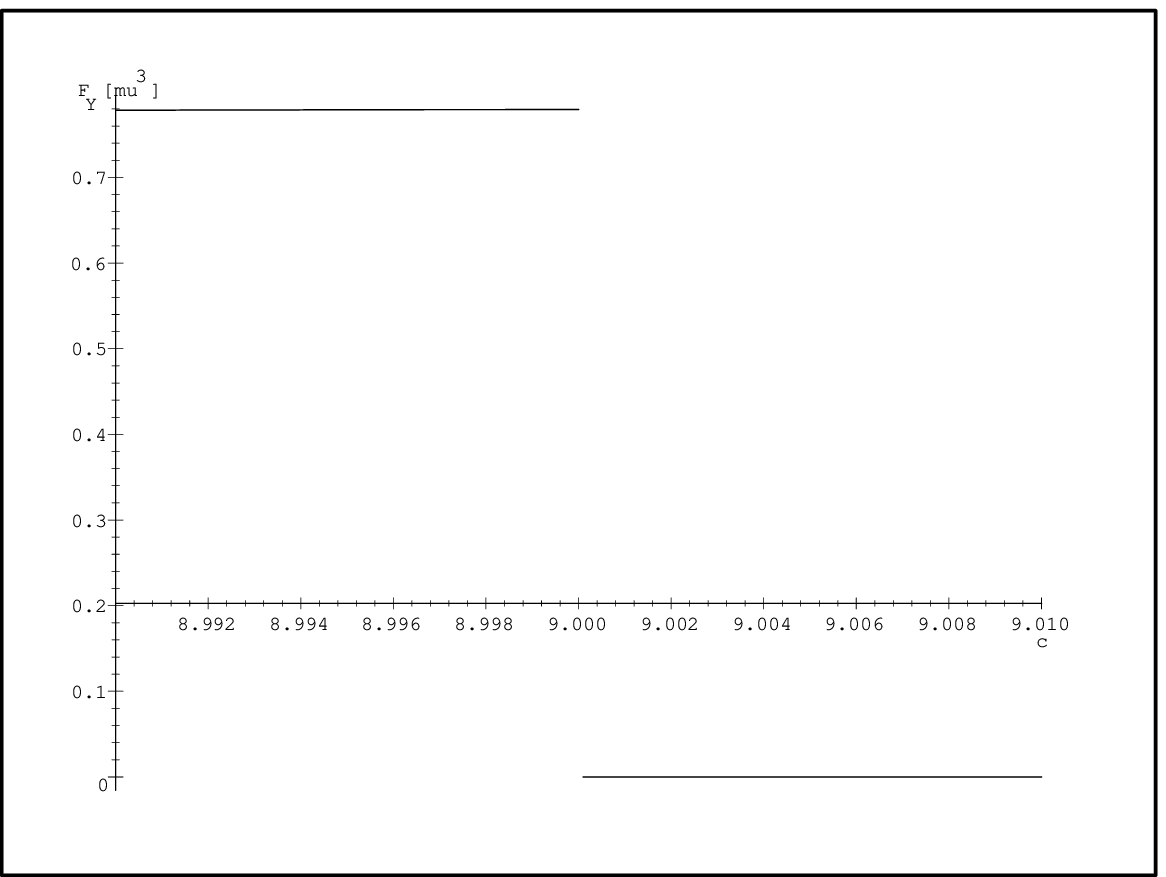}

{\small \em Fig. 4 - Scale of supersymmetry breakdown with respect to the
constant $c$}

\vspace{0.3cm}

Whereas models with only the condensate field are obviously not very
realistic\footnote[1]{As pointed out in \cite{nil1} it is difficult to transmit
SUSY-breaking to the matter sector in models whose hidden sector consists
exclusively of gaugino condensates, also, string inspired models generically
contain dilaton and moduli.},
these toy models can be studied to see which features are in principle
available for SUSY breaking scenarios. Particularly one would like to see a
realistic model containing a dilaton and moduli\footnote[2]{By moduli we mean
gauge singlet chiral fields which enter couplings of the effective Lagrangian,
but otherwise have no sources of potential -- like moduli in string inspired
models}, where supersymmetry is broken
(preferably by the auxiliary field of the condensate) and the cosmological
constant is 0. To determine what can be done to achieve these goals,
we proceed to
incorporate an additional field into our models, namely the dilaton, which
has to
be present in any string inspired supergravity model.

The way to incorporate the dilaton is to introduce a nontrivial gauge kinetic
function as the coefficient of $Y^3$ (formerly $U$) in the superpotential and
to add a suitable term to the K\"ahler function, exactly as in global
effective sigma models discussed in chapter 2, and then to put them into the
formula (\ref{potloc}). In what follows in this chapter we commit ourselves to
the no-scale form
$K=-\log(S+\bar{S})$ and $f=S$ as dictated by perturbative string
calculations\footnote[3]{However, the
general unpleasant features described below do not depend on the specific form
of the K\"ahler potential for the dilaton, for instance we could have
taken equally well $K=S \bar{S}$.}.

Incorporating the dilaton in this way into the models defined by
(\ref{k1}-\ref{w2}) makes
it clear that their local versions exhibit serious deficiencies. In the case of
global SUSY flat directions of vacua with unbroken supersymmetry exist in the
models.  Locally
supersymmetric versions are  even worse: the dilaton will either run to 0 or to
$\infty$,
resulting in a strongly coupled or free theory, neither of which is a viable
solution. This is a general feature of the locally supersymmetric potentials
associated with models
we discuss here when $f=S$, which is a consequence of the
symmetry (\ref{gsh}) of the superpotential.
It holds also for the non-minimal kinetic term (\ref{k2}).
Hence, the attractive possibility suggested at the end of chapter 2 does not
seem to be realized in the actual supergravity Lagrangian.

Incorporating additional moduli and matter fields does not
give attractive solutions as well: more than one gaugino condensate is needed
in any
case and the matter fields have to live in specially chosen representations
and have to acquire mass via their Yukawa couplings \cite{ccm}.
Of course, there is an important assumption which we make throughout
this investigation -- we demand a hierarchy between scales $\mu$, the
condensation scale and the Planck scale $M_{pl}$. However, the existence
of this hierarchy seems to be absolutely necessary if one wants to use
gaugino condensates for realistic phenomenology.

Considering the above results one is tempted to conclude
that the general conden\-sa\-te\--dilaton structure of the hidden sectors
discussed
so far is insufficient to create an acceptable minimum fixing in a stable way
values of all the fields involved (moduli and condensate) and breaking
supersymmetry. The last resort which is left is to modify the superpotential
for the dilaton, i.e. to take $f \neq S$. When one tries, it quickly becomes
obvious that not every modification would work. In this situation the best one
can do is to try to control the modifications through some new symmetry.
This is the approach we discuss in the next chapter.

\section{S-dual effective Lagrangians}

Here we discuss a viable solution to the problem of the run-away dilaton
(proposed in  \cite{apz1}):
taking a non-standard coupling of the condensate to the dilaton in the
superpotential.
Guessing naively, and  having in mind that a reasonable expectation
value of the dilaton
lies somewhere in the neighborhood of unity, one would postulate
a superpotential of the form $W \sim S+1/S$. Amazingly enough,
this simple superpotential falls close to the  superpotentials
realizing the principle of S-duality in the low energy effective Lagrangians.
Having amassed in the preceding chapters the evidence for the need of
modifications in the S-dependence of the effective superpotential,
we find that S-duality, discussed recently both in the context of strings
and in the context of N=1, N=2 supersymmetric Yang-Mills theories, is
the best motivated candidate to control the required corrections, and the one
which seems naturally suited to fulfill our  expectations.

As pointed out in \cite{apz1} there are two different nontrivial ways of
realizing the S-duality in the Lagrangians of the type we discuss here.
In its simplest realization, S-duality is an $SL(2,Z)$ symmetry generated by
$S \rightarrow 1/S, \; S\rightarrow S+i$. We shall discuss two physical
realizations of S-duality which differ in the way the gaugino sector
transforms under the action of the first generator:

\begin{description}
\item[Type-I S-duality:] here we assume that the gaugino sector closes under
the S-duality transformation. This states the invariance of $f Y^3$ under
S-duality. We have shown in \cite{apz1} that one can then redefine fields and
assume that $Y$ and $f$ are both independently invariant under the duality
transformation. Thus it is described by $S\rightarrow 1/S$ and $f
\rightarrow f$ (or equivalently $g^2 \rightarrow g^2$).

\item[Type-II S-duality:] if the gaugino sector (the `electric condensate')
does not close one has to take an additional sector (the `magnetic
condensate'). Only then one can have true strong-weak coupling duality. Type-II
S-duality is therefore defined by the condition $f\rightarrow 1/f$ (or
equivalently $g^2 \rightarrow 1/g^2$). Under the same transformation
these condensates would be interchanged.
\end{description}

\subsection*{Type-I S-duality}

Because we can assume that the gaugino condensate does not transform under this
S-duality \cite{apz1}, we are forced to consider a $SL(2,Z)$-invariant gauge
kinetic function $f$. Demanding that for large $S$ the $f$ should behave
asymptotically like $S$ (giving the old theory in the weak-coupling limit), we
take

\be
f=\frac 1{2\pi} \ln (j(S)-744),
\ee

where $j(S)$ denotes the usual generator of modular-invariant functions. Since
the superpotential must achieve a modular weight of $-1$ to cancel the
contributions of the K\"ahler potential, we are forced to include an
$\eta^2(S)$ prefactor into the superpotential. We choose

\be
K=-\ln(S+\Sb) +Y\Yb
\ee

and

\be
W=\frac {Y^3}{\eta^2(S)}
   (\frac 1{2\pi} \ln (j(S)-744) - 3 b\ln \frac Y\mu + c_0)+c,
\ee

where we have again included a constant $c$ parameterizing unknown effects
which
do not depend on $S$ and $Y$. Not surprisingly, this superpotential breaks
explicitly the ``accidental'' scale invariance of the superpotential
(\ref{scal}),  (\ref{gsh}).
Note that the  constant we include into the superpotential breaks S-duality,
therefore one can study whether the properties of the potential which are
created by S-duality are stable under perturbations.

For $c=0$ the
 scalar potential of this model exhibits a well defined minimum,
regardless whether one considers the SUGRA case or goes to the global SUSY
limit. Unfortunately, however, supersymmetry is unbroken in both cases.
Changing $c$ to a non-zero value does not help, either. Up to some critical
value of $c$, the minimum continues to exist and supersymmetry stays unbroken,
but for larger values of the constant the minimum becomes unstable and
vanishes.

We also studied the S-dual extension of the model given by (\ref{w2},
\ref{k2}), therefore taking non-minimal kinetic terms for the gaugino
condensate into account. Nevertheless we could not break supersymmetry
regardless of the value of the constant. It is also easy to see, why there is
a fundamental difference to the one-field model: the
scalar potential (\ref{potloc}) can be written as

\beq
V= e^K (g^{S\Sb} G_S G_{\Sb} + g^{Y\Yb}  G_Y G_{\Yb} -3 W \Wb)
\eeq

where $G_x=K_x W - W_x$. It is the term containing the metric $g^{Y\Yb}
=K_{Y\Yb}^{-1}$,
which is responsible for the singularity. But whereas $G_Y$ has been a function
of only $Y$ before, it is now a function of $S$ and $Y$. This additional
freedom allows one to find values for $S$, where $G_Y=0$ at the singularity in
$K_{Y\Yb}^{-1}$. Therefore paths exists, along which the vevs can slide around
the pole and then fall into the negative side of the pole. By a specific choice
of the constant $c_0$ (changing in effect the condensation scale $\mu$) it is
possible to avoid these zeroes of $G_Y$ at the singularity, thus confining the
vevs to the left side of the pole as in the $Y$-only model. But further
analysis shows  that in these cases supersymmetry is unbroken.

If adding the dilaton degree of freedom to a model destroys its supersymmetry
breaking properties, one would expect that adding further fields, i.e. the
generic modulus $T$ (the `breathing mode' common to all string
compactifications) does not change the picture any further.  But this is not
the case.

Adding a modulus $T$ in the usual $T$-duality invariant way \cite{venmag} gives
rise to a combined model with $S$- and $T$-duality, given by

\be
K=-\ln(S+\Sb) - 3 \ln (T+\Tb-Y\Yb),
\ee
and
\be
W= \frac {Y^3}{\eta^2(S) j(S)^{q/3}}
  (\frac 1{2\pi} \ln (j(S)-744) - 3 b\ln \frac Y\eta^2(T)\mu + c_0)+c.
\ee

This is actually a family of models because of the $j(S)^{q/3}$ factor in the
denominator of the prefactor. We include this factor because it does not change
the S-modular weight of $W$. In principle there could be any function of
$j(S)$, although one has to watch out for singularities. For $q>0$
one gets a scalar potential which vanishes for $S\rightarrow \infty$ at any
fixed value of $T$ and $Y$ (for a discussion of this property, see
\cite{apz1}).
Under $T$-duality we assume the condensate field $Y$ to transform as a modular
form of weight $-1$, so that the superpotential has correct modular weight -3
under $T$-duality (if $c=0$). The constant $c$ breaks both S- and T-dualities.
The second  constant $c_0$
 can in principle be adjusted to change the scale $\mu$ (it can be
reabsorbed into it \cite{apz1}). We adjust the value so that for $S=1$ and
$T=1$
the minimum of the scalar potential is at $Y=\mu$, thus setting the
condensation scale to $\mu^3$ in the case of $c=0$ and global SUSY (where the
actual minimum is at $S=1$, $T=1$).

This model exhibits SUSY breakdown with realistic expectation values of the
fields.
For $c=0$ (unbroken S- and T-duality) one finds a minimum at $S=1$, $T\simeq
1.23$, $Y\simeq \mu$. The SUSY breaking scale is determined by the expectation
value of the auxiliary field of the modulus $\expt{F_T}\simeq \mu^3$. This is
much larger than the other auxiliary fields: $\expt{F_Y} \simeq \mu \expt{F_T}$
and $\expt{F_S}=0$.

The cosmological constant is negative and of the order $-\expt{F_T}^2\simeq
-\mu^6$.

For small values of the constant ($c=O(10^{-17})$) this solution is still
stable, although several things happen with increasing $c$.

\bi
\i $F_S$ becomes nonzero and becomes larger when $c$ grows and reaches up to
$\simeq 10 \% $ of $F_T$,
\i $V_0$ increases with $c$ and eventually crosses value 0 and becomes
positive. Thus there is the possibility of cancelling the cosmological
constant.
\ei

For a larger value of $c$ the scalar potential becomes unstable and the dilaton
runs away to infinity. Fig. 5 shows roughly the qualitative behaviour of the
expectation values of the auxiliary fields and the value of the cosmological
constant with respect to $c$. For larger values of $c$ the theory becomes
unstable and the dilaton will run away, but in some intermediate region we can
make no statement on the existence of a minimum (searching
for a minimum is numerically very demanding, because the terms in the
scalar potential differ by up to 20 orders of magnitude).

\epsfbox[-80 -370 500 -100]{fg5.ps}

{\small \em Fig. 5 - Qualitative behaviour of auxiliary fields and cosmological
constant as a function of $c$}

\vspace{0.3cm}

Nevertheless one has for example a satisfactory model with
$q=1$, $c\simeq 7.3\,10^{-18}$, $\mu=10^{-5}$, where the vacuum is at
$S=1.002,\, T=1.234,\, Y=1.127 \mu$, which has a vanishing cosmological
constant and supersymmetry is broken: $\expt{F_T}=6\,10^{-18}$,
$\expt{F_S}=2\,10^{-20}$ and   $\expt{F_Y}=3\,10^{-23}$. Therefore the
principle of S-duality allows one to break supersymmetry without going
to multiple condensates or complicated matter representations, and with
a simple constant, parameterizing unknown and duality violating contributions,
one can also adjust the cosmological constant to 0.

One should take note of the fact, that contrary to our model with the gaugino
condensate only, the constant $c$ in this model is very small in comparison
to typical values of the rest of the superpotential. This makes this scenario
much more appealing, because we cannot and should not expect that the gaugino
superpotential is the only contribution (in principle these contributions could
come from the Yukawa sector of the superpotential, which we consider to be
non-existent throughout our analyses).

An interesting observation in itself is the statement, that if one considers a
symmetry acting on $S$ (in our case S-duality), this generically forces oneself
to postulate the full effective theory to all loops. If one postulates
a symmetry (acting on $S$) at the string tree-level, then at one loop this
symmetry will typically either be broken due to anomaly cancelling
contributions, or the anomaly will stay uncancelled.

Assume an effective supergravity theory at string tree-level. Under S-duality
$S \rightarrow (aS-ib)/(icS+d)$ the K\"ahler potential transforms as

\be
K \rightarrow K+\Phi(S)+\Phi(\bar S),
\ee

with $\Phi(S)=\ln(icS+d)$. To cancel this modular anomaly the gauge kinetic
function $f$ has to transform like \cite{dfkz}

\be
f/\eta^2(S) \rightarrow f/\eta^2(S)-a \Phi(S),
\ee

whereas it was designed to transform like

\be
f/\eta^2(S) \rightarrow f/\eta^2(S) (icS+d).
\ee

To achieve this, $S$ has to transform in a very complicated way. This will
break
S-duality and even after changing the K\"ahler function like in \cite{dfkz},
the new Lagrangian can hardly be invariant under the $SL(2,Z)$ transformation
on $S$ any longer while still having the same K\"ahler transformation $\Phi(S)$
under the symmetry.

This seems to be a general principle: if the theory is at tree-level invariant
under a symmetry acting on $S$ which produces a K\"ahler transformation
(depending on the dilaton)
$\Phi(S)$, then adding the anomaly cancelling terms generically makes it
impossible to change the K\"ahler function and superpotential in such a way,
that

\begin{enumerate}
\item the theory is still invariant under the S-transformation, and that
\item the K\"ahler transformation $\Phi(S)_{1-loop}$ associated with the old
symmetry
and the modified K\"ahler function is still equal to $\Phi(S)_{tree}$. This has
to be the case, because otherwise the theory would not be anomaly free,
because the counter\-terms designed to cancel the anomaly coming from
$\Phi(S)_{tree}$ will not cancel the one coming from $\Phi(S)_{1-loop}$.
\end{enumerate}

Thus it seems not to be possible to promote a specific symmetry on $S$ to
higher loops. On the contrary, if one proposes a symmetry like S-duality on the
level of the fundamental string theory, this symmetry cannot be present in an
effective action unless it is an action to all orders (if it is associated with
a K\"ahler transformation). Therefore, since we
want to impose S-duality as a physical
principle, we have to demand the $SL(2,Z)_S$ invariance at the level of the
full theory to all loops including non-perturbative effects.

Of course this shows, that our model can only be a toy model to show that
imposing symmetry constraints can solve the problems of gaugino
condensation. In a more realistic example we would expect to have mixing
between the dilaton and the modulus, since we know that these are
present already at the 1-loop level due to target-space modular invariance
\cite{lm}.

\subsection*{Type-II S-duality}

The type-II implementation of S-duality takes into account the fact that the
gaugino sector of the theory might not close under the S-duality
transformation. To write down a model which is invariant one has to include an
additional sector, the `magnetic condensate', which is supposed to represent
the dual phase of the theory \cite{sw}.

The simplest toy model which illustrates the idea of type-II S-duality is given
by

\beq
K = -\ln (S+\bar S) - 3 \ln (T+\bar T-Y \bar Y - H \bar H),
\eeq

\beq
W= \frac 1{\eta^2(S)} (Y^3 S + H^3/S + 3 b Y^3 \ln \frac{Y \eta^2(T)}{\mu}
+ 3 b H^3 \ln \frac{H \eta^2(T)}{\mu}+Y^3 H^3/\mu^3).
\eeq
This model does not exhibit a full $SL(2,Z)$-symmetry, but only the
strong-weak-coupling duality

\beq
f=S \rightarrow 1/S, \qquad Y \leftrightarrow H.
\eeq
In principle one could with some effort promote this symmetry to a full
$SL(2,Z)$, but for illustration of our statements we choose this simple model,
especially because both real and imaginary parts of $S$ already become fixed
even with this smaller symmetry.

Again, the unwanted scale symmetry (\ref{scal}), (\ref{gsh}) is not realized
in the superpotential.
The scalar potential possesses a (although rather hard to find) minimum close
to
$S=1, T=0.560, H=Y=\mu^\prime= 3.64\,10^{-2}\, \mu $\cite{apz1}. It turns out
that at the
minimum supersymmetry is broken, with the magnitude of SUSY breaking again
determined by $\expt{F_T} \simeq {\mu^\prime}^3/M$, where $\mu^\prime$ is
the dynamically determined value of the condensate at the minimum
(in the previous type-I examples we adjusted (using $c$) $\mu^\prime$ to be of
the phenomenologically reasonable value $10^{-5}$). The cosmological constant
is again negative and of the order $\simeq -{\mu^\prime}^6$.

\section{General aspects of supersymmetry breaking}

All our models exhibit one universal feature, which
seems to hold in all gaugino condensation models with dilaton and modulus,
regardless
of whether one works in the effective Lagrangian or the effective
superpotential approach: it is always $F_T$ which achieves the dominant
vacuum expectation value of all the auxiliary fields. Of course one would like
to know, whether this feature is generic or not. It should be made clear,
that most of the models are constructed in a way which makes $\expt{F_S}$ small
or zero. In our models of the previous chapter we have assumed S-duality,
therefore the scalar potential will always have an extremum at $S=1$ and in our
models this happens to be a minimum. But one can easily calculate that (if
$c=0$) $F_S(S=1)=0$.  We showed that breaking the S-duality (by taking a
constant into the superpotential) increased the value of $F_S$.  So we believe
that the smallness of $F_S$ is caused by our specific construction. The same is
true in models where multiple gaugino condensates are used to fix the vev of
the dilaton. These fix the value
of the modulus $T$ by looking at only the minima which correspond to minima of
the tree-level approximation. These are guaranteed to have $\expt{F_S}=0$ due
to the fact that the superpotential factorizes into $S$ and $T$ dependent parts
\cite{ccm}.  One-loop corrections are then able to give
$F_S$ contributions, but these can be expected to be small. But it is not
clear, whether these corrections do not introduce new minima which have
$\expt{F_S}$ of the same order as $\expt{F_T}$.

This shows the possibility that $\expt{F_S}$ could in principle become the
dominant supersymmetry breaking contribution, and that statements which claim
that $F_S$ is shown to stay small generically should be taken carefully.

Finally, let us examine the mass hierarchy of different terms in the potential
of a typical condensate-moduli model \cite{kl}. The superpotential of these
models
naturally appears in such a form, that the magnitude of the superpotential at
the SUSY breaking minimum (if there is any) is of the order $\mu^3$
(condensation scale cubed) which means that the gravitino mass term which
usually sets the magnitude of the soft breaking terms in the observable sector
is of the order of $\kappa^2 \mu^3$, i.e. lies in the TeV range when $\mu$ is
of the order of a typical condensation scale for unification groups.  When one
writes down the potential and groups together terms of different order in $\mu$
one gets (after careful restitution of powers of the Planck scale)

\beqa V_1 =
g^{U \bar{U}} W_U \bar{W}_{\bar{W}} &\sim& \mu^4 \\
V_2 = \kappa^2 g^{U
  \bar{U}} K_U W \bar{W}_{\bar{U}} + \; h.c. & \sim& \kappa^2 \mu^6
\label{k2m6} \\
V_3 = \kappa^2 (S+\bar{S})^2 e^{\kappa^2 K}
|-\frac{1}{S+\bar{S}} W +W_S|^2 &\sim& \kappa^2 \mu^6 \\
V_4 = - 3 \kappa^2
|W|^2 &\sim& \kappa^2 \mu^6 \\
V_5 = \kappa^4 g^{U \bar{U}} |K_U W |^2 &\sim&
\kappa^4 \mu^8 \label{k4m8} \eeqa

Among these terms the first one is dominating
the potential, so one could try to find the zeroth-order minimum just
minimizing this dominant term.  The condition which is used widely to determine
this approximate minimum is $W_U = 0$ which in turn allows, if the
superpotential is simple enough, to express the condensate through the dilaton
in terms of superfields.
Then one introduces this relation into the original superpotential obtaining
an effective superpotential for $S$ only, which is then processed in a usual
way in quest for the minimum in $S$. In simple models this approach, known as
effective superpotential method, gives a pretty correct description of the
behaviour of $S$, but one has to be rather careful in drawing conclusions this
way in more sophisticated models. First, the `new' effective potential misses
the terms (\ref{k2m6}) and (\ref{k4m8}), the first of which is formally of the
order of
the terms which are left. Second, the actual condition which enforces the
unbroken global supersymmetry is $g^{U \bar{U}} \bar{W}_{\bar{U}} =0$ which in
a case of a general K\"ahler potential for the condensate may have several
solutions, some of them corresponding to vanishing of $g^{U \bar{U}}$.  Also,
in general discussions of the cosmology of potentials induced by gaugino
condensation one should study the full model. The condition of the type
$\partial W / \partial U =0$ singles out a curved `valley' in the full
potential, and in general excitations orthogonal to the valley are possible in
the early universe as is the existence of other, disconnected valleys. Finally,
as was demonstrated by our examples with constant terms in the superpotential,
the
existence of terms which are normally irrelevant for global supersymmetry can
change dramatically the vacuum properties of local models, effectively changing
the above discussed hierarchy of terms in the potential, thus invalidating the
integrating-out procedure based on the condition $\partial W / \partial U =0$.
Indeed, even naive counting of powers of mass scales shows that if there is a
constant of the order of $\mu / \kappa^2$ in the superpotential, the omitted
terms (\ref{k2m6}) and (\ref{k4m8}) are in fact of the order of $\mu^4$, i.e.
as important as the would-be leading term (although it should not be forgotten,
that the constant which we can use to adjust the cosmological constant to 0 is
much smaller: $c\simeq 0.01\;\mu^3$).  Finally, we have shown for
type-II S-duality in \cite{apz1}, that in cases where condensates couple to
different  functions of
the dilaton, like $S$ vs $1/S$, solving the simple integrating-out conditions
becomes ambiguous if possible at all.

\section{Discussion and conclusions}

In the present paper we have analyzed the effective
Lagrangian pertinent to supergravity hidden sectors composed of gauge fields,
dilaton and other moduli.

We tried to construct models which are phenomenologically as realistic as
possible, demanding that they possess potentials with
stable minima corresponding to reasonable expectation values of all fields and
exhibit supersymmetry broken so as to produce an acceptable gravitino mass in
the TeV range.

We have been searching for such minima both analytically and using
high precision numerical methods. In pure gauge models, without dilaton, we
have found that the breaking of supersymmetry requires introduction of
non-symmetric (non-minimal) K\"ahler functions or/and adding a sufficiently
large constant to the superpotential (the M-term) in the local case.  When one
adds the dilaton to these models, the situation becomes much more
difficult. With
only the simplest linear coupling of the dilaton to the gaugino condensate we
were unable to find a model with a well defined nontrivial minimum, even
including an arbitrary constant into superpotential and taking a non-symmetric
K\"ahler function for the condensate. Also, we have confirmed that the global
models which without the dilaton featured nontrivial (although supersymmetric)
minima become ``ill'' when one couples to them the dilaton in a linear way -
they
generally acquire a flat or runaway direction associated with dilaton.
Motivated by this evidence we suggest the need to modify the coupling of the
dilaton to the gauge sector.  We have proposed and discussed in detail
modifications of the dilaton-gaugino effective Lagrangian consistent with the
principle of S-duality.  We have identified two different, physically
nontrivial, ways of incorporating S-duality into the low-energy effective
Lagrangian, which differ in the way the coupling constant is transformed. Both
implementations lead easily to physically satisfying solutions, i.e. stable
minima with broken supersymmetry, while employing only a single gaugino
condensate and without the need for specially constructed matter sectors.  We
feel that this solution to the long-standing problem of runaway
dilaton vacua is more attractive than the traditional methods relying on
multiple condensates. In general,
although we cannot prove this in a mathematically rigorous way, it seems
reasonably
safe to conclude that in the full nonperturbative effective Lagrangian coming
from superstring the coupling of the dilaton is likely to be significantly
different from the simple linear one implied by perturbative calculations if a
hidden gauge sector is the source of supersymmetry breaking and mass hierarchy.

\subsection*{Acknowledgments}
This work was supported by the Deutsche Forschungsgemeinschaft and EC grants
SC1--CT92--0789 and SC1--CT91--0729. Z.L. has been supported by the A. von
Humboldt Fellowship. A. N. has been supported by a PhD scholarship from the
Technical University of Munich. We would like to thank Andre Lukas, Peter
Mayr and Dimitris Matalliotakis for useful discussions.


\begin{thebibliography}{000}

\bibitem{rev} H.P. Nilles, Phys.Rep. {\bf 110} (1984) 1;
G.G. Ross, {\em Grand Unified Theories}, Addison-Wesley Pub. 1985;
S. Ferrara (ed), {\em Supersymmetry}, 2 vols, North-Holland 1987.
\bibitem{nil1}  H.P. Nilles, Phys.Rev.Lett. {\bf 115B} (1982) 193;
\bibitem{nfer}   S. Ferrara, L.
Girardello, H.P. Nilles, Phys.Lett. {\bf 125B} (1983) 457;
\bibitem{nrev}  J.P. Derendinger,
L.E. Ib\'a\~nez, H.P. Nilles, Phys.Lett. {\bf 155B} (1985) 467; M. Dine,
R. Rohm, N. Seiberg, E. Witten, Phys.Lett. {\bf 156B} (1985) 55; T.R. Taylor,
Phys.Lett. {\bf 164B} (1985) 43.
\bibitem{ccm} N.V. Krasnikov, Phys.Lett. {\bf 193B} (1987) 37; J.A. Casas,
Z. Lalak, C. Mu\~noz, G.G. Ross, Nucl.Phys. {\bf B347} (1990) 243; for an
extended list of references see: B. de Carlos, J.A. Casas, C. Mu\~noz,
Nucl.Phys. {\bf B399} (1993) 623.
\bibitem{wit} E. Witten, PHys.Lett. {\bf 155B} (1985) 151
\bibitem{witsen}
J. Schwarz, A. Sen, Nucl.Phys. {\bf B411} (1994) 35; for an extended list of
references see:
A. Sen, Int.J.Mod.Phys {\bf A9} (1994) 3707.
\bibitem{fi} A. Font, L. Ib\'a\~nez, D. L\"ust, F. Quevedo, Phys.Lett. {\bf
249B} (1990) 35;
J. Schwarz, A. Sen, Nucl.Phys. {\bf B411} (1994) 35.
\bibitem{hm} J.H. Horne, G. Moore, ``{\em Chaotic Coupling Constants}'',
hep-th/9403058.
\bibitem{veny} G. Veneziano, S. Yankielowicz, Phys.Lett. {\bf 113B} (1982)
231.
\bibitem{ferzum} S. Ferrara,B. Zumino, Nucl.Phys. {\bf B 79} (1974) 413.
\bibitem{witt} E. Witten, Nucl.Phys {\bf B 202} (1982) 253.
\bibitem{rossi} D. Amati, K.Konishi, Y.Meurice, G.C.Rossi, G.Veneziano,
Phys.Rept. {\bf 162} (1988) 169.
\bibitem{wesov} B. Ovrut, J. Wess, Phys.Rev. {\bf D 25} (1982) 409.
\bibitem{venmag}  S. Ferrara, N. Magnoli, T.R. Taylor, G. Veneziano,
Phys.Lett. {\bf 245B} (1990) 409, A. Font, L.E. Ib\'a\~nez, D. L\"ust,
F. Quevedo,
Phys.Lett. {\bf 245B} (1990) 401; H.P. Nilles, M. Olechowski, Phys.Lett. {\bf
248B} 1990 268; P. Binetruy, M.K. Gaillard, Phys.Lett {\bf 253B} (1991) 119;
M. Cvetic, A. Font, L.E. Ib\'a\~nez, D. L\"ust, F. Quevedo,  Nucl.Phys. {\bf
B361} (1991) 194.
\bibitem{tayrey} S.-J. Rey, T.R. Taylor, Phys.Rev.Lett {\bf 71} (1993)
1132.
\bibitem{apz1} Z. Lalak, A. Niemeyer, H.P. Nilles, prep. TUM-HEP-202/1994,
Phys. Lett. {\bf B} in press.
\bibitem{dfkz} J.-P. Derendinger, S. Ferrara, C. Kounnas, F. Zwirner,
Nucl. Phys. {\bf B372} (1992) 145.
\bibitem{lm} D. L\"ust, C. Mu\~noz, Phys.Lett. {\bf 279B} (1992) 272
\bibitem{sw}   C. Montonen, D. Olive,
Phys.Lett. {\bf 72B} (1977) 117; C. Vafa, E. Witten, ``{\em A strong coupling
test of S-duality''}, hep-th/9408074; N. Seiberg, E.Witten, hep-th/9411149;
K. Intrilligator, N. Seiberg, hep-th/9408155, E. Witten, hep-th/9503124
\bibitem{kl} V. Kaplunovsky, J. Louis, Nucl.Phys. {\bf B422} (1994) 57
\end{thebibliography}
\end{document}